\pdfoutput=1
\documentclass[lettersize,journal]{IEEEtran}
\usepackage{amsmath,amsfonts}
\usepackage{array}
\usepackage{textcomp}
\usepackage{url}
\usepackage{graphicx}
\usepackage{cite}
\usepackage{booktabs}
\usepackage{multirow}
\usepackage{algorithmic}
\usepackage{algorithm}
\usepackage{tikz}
\usetikzlibrary{arrows.meta,positioning,decorations.pathreplacing}

\begin{document}
\title{Graduated Trust Gating for IoT Location\\
Verification: Trading Off Detection\\
and Proof Escalation}

\author{%
Yoshiyuki Ootani%
\thanks{Yoshiyuki Ootani is an independent researcher based in Japan (e-mail: info@ootanl.com).}%
}

\maketitle

\begin{abstract}
IoT location services accept client-reported GPS coordinates at
face value, yet spoofing is trivial with consumer-grade tools.
Existing spoofing detectors output a binary decision, forcing
system designers to choose between high false-deny and high
false-accept rates. We propose a graduated trust gate that
computes a multi-signal integrity score and maps it to three
actions---\textsc{proceed}, \textsc{step-up}, or
\textsc{deny}---where \textsc{step-up} invokes a stronger
verifier such as a zero-knowledge proximity proof. A session-latch
mechanism ensures that a single suspicious fix blocks the entire
session, preventing post-transition score recovery. Under an
idealized step-up oracle on 10\,000 synthetic traces, the gate
enables strict thresholds ($\theta_p = 0.9$) that a binary gate
cannot safely use: at matched false-accept rate (11\%), the
graduated gate maintains zero false-deny rate versus 0.05\% for
binary, with 5\,$\mu$s scoring overhead. Real-device traces from
an Android smartphone demonstrate the session-latch mechanism and
show that a nearby mock location ($\sim$550\,m) evades
$\theta_p = 0.7$ but is routed to step-up at $\theta_p = 0.9$.
Signal ablation identifies a minimal two-signal configuration
(F1\,=\,0.84) suitable for resource-constrained scoring layers.
\end{abstract}
\begin{center}
\small This work has been submitted to the IEEE for possible publication.
Copyright may be transferred without notice, after which this version may no longer be accessible.
\end{center}

\section{Introduction}
\label{sec:intro}

\IEEEPARstart{L}{ocation-dependent} IoT services---geofenced access
control, location-bound content, and proximity-triggered
actions---rely on client-reported GPS coordinates that are trivially
spoofable via mock-location APIs~\cite{mock-gps} or software-defined
radio~\cite{sdr-spoofing}.

Spoofing detection techniques (sensor fusion~\cite{guardian-gps},
RAIM-inspired consistency~\cite{raim-original}, ML
classifiers~\cite{ml-spoof-detect}) produce a binary decision but
do not prescribe a \emph{response}. A binary accept/deny gate forces
a hard trade-off: a strict threshold causes false denials; a lenient
one lets spoofers through.

We propose \emph{graduated trust gating}: a three-level decision
layer that converts an \emph{imperfect lightweight detector} into a
practical access-control mechanism. The gate maps a trust score $T$
to \textsc{proceed} ($T \ge \theta_p$), \textsc{step-up}
($\theta_s \le T < \theta_p$), or \textsc{deny} ($T < \theta_s$).
\textsc{Step-up} triggers a Groth16 zero-knowledge proximity
check only in the sense of \emph{escalation}: our evaluation treats
step-up as a stronger verifier, not as a fully instantiated trust
source for location truth. The key insight is that the gate does not need
a perfect detector---it only needs to separate high-confidence
legitimate traffic from the rest, then delegate uncertain cases to a
stronger verifier.

\smallskip\noindent\textbf{Contributions.}
(C1)~A graduated gate design that provides a tunable FAR--friction
trade-off via a single threshold $\theta_p$
(\S\ref{sec:design}).
(C2)~Signal ablation identifying a minimal two-signal configuration
(temporal consistency + network cross-check, F1\,=\,0.84) suitable
for resource-constrained IoT endpoints (\S\ref{sec:eval:ablation}).
(C3)~Robustness analysis under six signal-degradation scenarios,
showing that legitimate traffic remains above the proceed threshold
in all cases (\S\ref{sec:eval:robustness}).

\smallskip\noindent
We use \emph{V1} to denote the baseline three-signal scorer
(S1--S3) and \emph{V2} for the full five-signal scorer (S1--S5).

\section{Related Work}
\label{sec:related}

\textbf{Spoofing detection.}
Sensor-fusion approaches cross-check GPS against IMU, Wi-Fi, or
cell-tower data~\cite{guardian-gps}. RAIM adapts satellite integrity
monitoring to consumer devices~\cite{raim-original}. ML classifiers
train on labelled trajectories~\cite{ml-spoof-detect}. These methods
primarily output binary trust decisions; our focus is the response
layer that follows from such imperfect detectors.

\textbf{Location proofs.}
Groth16 zero-knowledge proofs~\cite{groth16} enable proximity
verification without coordinate disclosure. FibRace~\cite{fibrace}
benchmarks client-side proving across 6{,}000 mobile devices,
showing that mobile proving is feasible but highly device-dependent.
Our gate invokes such proofs \emph{selectively}, reducing
average-case cost.

\textbf{Access control for LBS.}
Prior work on location-based access control focuses on policy
languages and geofence evaluation. To our knowledge, no existing
system combines a multi-signal trust score with a three-level response
gate, session latch, and proof escalation.

\section{System Design}
\label{sec:design}

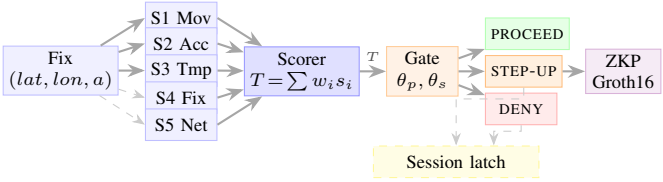
\begin{figure}[!t]
\centering
\begin{tikzpicture}[
  >=Stealth, node distance=0.3cm,
  bx/.style={draw, rectangle, minimum height=0.42cm, minimum width=0.95cm,
             font=\scriptsize, align=center, inner sep=2pt},
  sig/.style={bx, fill=blue!6, draw=blue!25},
  sc/.style={bx, fill=blue!12, draw=blue!40, minimum width=1.15cm, minimum height=0.62cm},
  gt/.style={bx, fill=orange!10, draw=orange!40, minimum width=0.95cm, minimum height=0.62cm},
  pro/.style={bx, fill=green!10, draw=green!30},
  ste/.style={bx, fill=orange!12, draw=orange!50},
  den/.style={bx, fill=red!8, draw=red!30},
  prf/.style={bx, fill=violet!8, draw=violet!30, minimum width=1.0cm},
  lat/.style={bx, fill=yellow!12, draw=yellow!50, dashed, minimum width=2.2cm},
  arr/.style={->, thick, draw=gray!70},
  arrd/.style={->, thin, draw=gray!40, dashed},
  lb/.style={font=\tiny, text=gray!60!black},
]
\node[sig] (inp) {Fix\\$(lat,lon,a)$};
\node[sig, right=0.35cm of inp, yshift=0.7cm] (s1) {S1 Mov};
\node[sig, right=0.35cm of inp, yshift=0.36cm] (s2) {S2 Acc};
\node[sig, right=0.35cm of inp, yshift=0.00cm] (s3) {S3 Tmp};
\node[sig, right=0.35cm of inp, yshift=-0.36cm] (s4) {S4 Fix};
\node[sig, right=0.35cm of inp, yshift=-0.72cm] (s5) {S5 Net};
\draw[arr] (inp) -- (s1.west);
\draw[arr] (inp) -- (s2.west);
\draw[arr] (inp) -- (s3.west);
\draw[arrd] (inp) -- (s4.west);
\draw[arrd] (inp) -- (s5.west);
\node[sc, right=0.35cm of s3, yshift=-0.01cm] (scorer) {Scorer\\$T\!=\!\sum w_i s_i$};
\draw[arr] (s1.east) -- (scorer);
\draw[arr] (s2.east) -- (scorer);
\draw[arr] (s3.east) -- (scorer);
\draw[arr] (s4.east) -- (scorer);
\draw[arr] (s5.east) -- (scorer);
\node[gt, right=0.35cm of scorer] (gate) {Gate\\$\theta_p,\theta_s$};
\draw[arr] (scorer) -- node[above, lb] {$T$} (gate);
\node[pro, right=0.35cm of gate, yshift=0.5cm] (pro) {\textsc{proceed}};
\node[ste, right=0.35cm of gate] (ste) {\textsc{step-up}};
\node[den, right=0.35cm of gate, yshift=-0.5cm] (den) {\textsc{deny}};
\draw[arr] (gate) -- (pro);
\draw[arr] (gate) -- (ste);
\draw[arr] (gate) -- (den);
\node[prf, right=0.3cm of ste] (zkp) {ZKP\\Groth16};
\draw[arr] (ste) -- (zkp);
\node[lat, below=0.65cm of gate, xshift=0.45cm] (latch) {Session latch};
\draw[arrd] (ste.south) -- ++(0,-0.15) -| (latch.north);
\draw[arrd] (den.south) -- ++(0,-0.08) -| ([xshift=0.5cm]latch.north);
\end{tikzpicture}
\caption{Graduated trust gating pipeline. Dashed arrows indicate
optional signals. Once \textsc{step-up} or \textsc{deny} fires,
the session latches.}
\label{fig:arch}
\end{figure}

Fig.~\ref{fig:arch} shows the end-to-end pipeline.

\subsection{Trust Score}

We compute $T = \sum_{i \in \mathcal{A}} w_i \, s_i$ where each
signal $s_i \in [0,1]$ and weights sum to~1. Five signals are
defined:

\smallskip\noindent
\textbf{S1} (Movement): speed between consecutive points; penalizes
$v > 50$\,m/s.
\textbf{S2} (Accuracy): flags $< 2$\,m (common in GPS simulators).
\textbf{S3} (Temporal): counts ``teleportation'' violations
($> 100$\,m/s) across history pairs.
\textbf{S4} (Fix consistency): RAIM-inspired scatter-vs-accuracy
ratio of recent raw GPS fixes.
\textbf{S5} (Network): distance between GPS position and cell/Wi-Fi
hint relative to hint accuracy.

When some signals are unavailable, weights follow the pre-defined
profiles in Table~\ref{tab:weights}.

\begin{table}[!t]
\caption{Weight Profiles by Available Signals\label{tab:weights}}
\centering\footnotesize
\begin{tabular}{lccccc}
\toprule
\textbf{Profile} & $w_1$ & $w_2$ & $w_3$ & $w_4$ & $w_5$ \\
\midrule
All five    & .30 & .10 & .15 & .25 & .20 \\
No network  & .35 & .15 & .20 & .30 & --- \\
No fixes    & .40 & .15 & .20 & --- & .25 \\
V1 (S1--S3) & .50 & .20 & .30 & --- & --- \\
\bottomrule
\end{tabular}
\end{table}

\subsection{Graduated Gate}

The gate maps $T$ to one of three actions:
\begin{equation}
G(T) = \begin{cases}
\textsc{proceed} & T \geq \theta_p \\
\textsc{step-up} & \theta_s \leq T < \theta_p \\
\textsc{deny}    & T < \theta_s
\end{cases}
\label{eq:gate}
\end{equation}

Default: $\theta_p = 0.7$, $\theta_s = 0.3$. The key parameter is
$\theta_p$: raising it routes more traffic through step-up,
reducing FAR at the cost of additional proof overhead. The gate does
not require a high-accuracy detector; it only needs the detector to
assign legitimate traffic consistently above $\theta_p$.

\subsection{Step-Up Verification}

When \textsc{step-up} is triggered, the client generates a Groth16
proximity proof~\cite{groth16} that, in our prototype, uses 474
constraints on the BN128 curve and proves that
it is within radius~$R$ of the target. However, such a proof is only
as strong as the trustworthiness of the location witness. A deployable
step-up scheme therefore requires trusted location evidence, such as
signed location tokens, proximity beacons, or TEE-backed measurements;
our evaluation abstracts this step as an oracle and uses Groth16 only
as a representative escalation primitive. On endpoints without step-up
capability, the system falls back to \textsc{deny}.

\subsection{Session Latch}

A per-fix gate evaluates each location report independently. However,
a spoofer who triggers one suspicious fix (e.g., teleportation) may
subsequently settle at the spoofed location with high scores. To
prevent this, we add \emph{session-latch semantics}: once the gate
transitions to \textsc{step-up} or \textsc{deny}, the session is
\emph{latched}---all subsequent fixes in the same session return the
latched state regardless of score. A \textsc{step-up} latch is cleared
only after successful external verification; a \textsc{deny} latch
persists until the session is restarted. This ensures that a single
suspicious transition blocks the entire session, not just one fix.

\smallskip
Algorithm~\ref{alg:gate} summarises the session-aware gate logic.

\begin{algorithm}[!t]
\caption{Session-aware graduated gate}\label{alg:gate}
\begin{algorithmic}
\STATE \textbf{Input:} fix $p$, history $H$, context $C$, session state $\ell$
\STATE \textbf{Output:} action $\in \{\text{proceed}, \text{step-up}, \text{deny}\}$
\IF{$\ell \neq \text{null}$}
  \RETURN $\ell$ \COMMENT{step-up: cleared by verification; deny: cleared by restart}
\ENDIF
\STATE $T \gets \sum_{i \in \mathcal{A}} w_i \cdot s_i(p, H, C)$
\IF{$T \geq \theta_p$}
  \RETURN proceed
\ELSIF{$T \geq \theta_s$}
  \STATE $\ell \gets \text{step-up}$ \COMMENT{latch session}
  \RETURN step-up
\ELSE
  \STATE $\ell \gets \text{deny}$ \COMMENT{latch session}
  \RETURN deny
\ENDIF
\end{algorithmic}
\end{algorithm}

\section{Evaluation}
\label{sec:eval}

\subsection{Setup}

\textbf{Synthetic traces.}
We generate 10\,000 synthetic traces (1\,000 per scenario, seeded
PRNG) across four legitimate (walking, driving, stationary, train)
and six spoofed scenarios. Code is open-source~\cite{zairn2026}.

\textbf{Real-device traces.}
We additionally collect traces from a Jelly Star smartphone
(Android~10, Chrome~145) under three conditions: (i)~honest walking
(30~fixes, 32\,s), (ii)~honest stationary (27~fixes, 30\,s), and
(iii)~mock-location teleportation via Android developer options
(58~fixes: 27 real GPS in Tokyo followed by 31 mock fixes at
Miami, FL---a $\sim$10{,}000\,km jump with accuracy 0.01\,m), and
(iv)~nearby mock location ($\sim$550\,m from true position,
61~fixes, accuracy 0.01\,m).

\subsection{Detection Accuracy and AUC}

Table~\ref{tab:scores} reports per-scenario mean trust scores and
overall classification metrics. V2 improves AUC-PR from 0.71 (V1)
to 0.93, and reduces equal-error rate from 0.20 to 0.08. Three
spoofed scenarios (drift, accuracy, net mismatch) score above
$\theta_p = 0.7$, evading heuristic detection. This motivates the
graduated gate: rather than demanding a perfect detector, the gate
routes uncertain cases to step-up verification.

\begin{table}[!t]
\caption{Mean Trust Score and Classification Summary\label{tab:scores}}
\centering\footnotesize
\begin{tabular}{llc|llc}
\toprule
\textbf{L} & \textbf{Scenario} & $\bar{T}$ &
\textbf{L} & \textbf{Scenario} & $\bar{T}$ \\
\midrule
Leg. & Walking    & 1.00 & Sp. & Teleport. & 0.50 \\
Leg. & Driving    & 1.00 & Sp. & Drift     & 0.94 \\
Leg. & Stationary & 0.99 & Sp. & Accuracy  & 0.73 \\
Leg. & Train      & 0.99 & Sp. & Replay    & 0.55 \\
     &            &      & Sp. & Net mism. & 0.86 \\
     &            &      & Sp. & Compound  & 0.59 \\
\midrule
\multicolumn{3}{c|}{V1: AUC-PR = 0.71, EER = 0.20} &
\multicolumn{3}{c}{V2: AUC-PR = 0.93, EER = 0.08} \\
\bottomrule
\end{tabular}
\end{table}

Table~\ref{tab:dist} shows the score distribution for legitimate
and spoofed classes, confirming clear separation: all legitimate
scores exceed $\theta_p = 0.7$, while spoofed scores span a wide
range with three scenarios above the threshold.

\begin{table}[!t]
\caption{Trust Score Distribution (V2, $N$\,=\,10{,}000)\label{tab:dist}}
\centering\footnotesize
\begin{tabular}{lcccc}
\toprule
\textbf{Class} & \textbf{Mean} & \textbf{Min} & \textbf{P25} & \textbf{Max} \\
\midrule
Legitimate (4{,}000) & 0.995 & 0.880 & 0.990 & 1.000 \\
Spoofed (6{,}000)    & 0.696 & 0.220 & 0.550 & 1.000 \\
\bottomrule
\end{tabular}
\end{table}

\subsection{Signal Ablation}
\label{sec:eval:ablation}

To isolate each signal's contribution, we evaluate all 31 non-empty
signal subsets with \emph{proportionally redistributed} weights
(i.e., disabled signals' weights are spread to the remaining signals,
unlike the predefined profiles in Table~\ref{tab:weights} used in
deployment). Under this ablation-specific weighting, the best minimal
configuration is S3+S5 (F1\,=\,0.84)---higher than all five signals
(F1\,=\,0.67), consistent with Table~\ref{tab:robust}. Note that S3's Shapley $\Delta$F1 is negative
($-$0.040) because its \emph{marginal contribution in the full-set
context} differs from its \emph{synergy in a sparse subset}: S3
adds little when S1 already covers movement anomalies, but becomes
essential when paired only with S5.
More broadly, signals such as S2 (accuracy) produce high scores even
for spoofed traces where accuracy is within normal range, diluting
the composite. In a weighted sum, a signal that cannot distinguish a
particular attack mode effectively \emph{votes for legitimacy},
pulling the composite above the threshold.
Sparse but complementary signals avoid this dilution: S3 catches
temporal anomalies while S5 catches network-position mismatches,
covering distinct attack surfaces without mutual interference.

\begin{table}[!t]
\caption{Signal Importance ($\Delta$F1, Shapley-like)\label{tab:ablation}}
\centering\footnotesize
\begin{tabular}{llc}
\toprule
\textbf{Signal} & \textbf{Description} & $\Delta$\textbf{F1} \\
\midrule
S5 & Network cross-check  & $+$0.103 \\
S4 & Fix consistency (RAIM) & $+$0.082 \\
S1 & Movement plausibility & $+$0.019 \\
S2 & Accuracy anomaly      & $-$0.019 \\
S3 & Temporal consistency   & $-$0.040 \\
\midrule
\multicolumn{2}{l}{Best two-signal: S3$+$S5} & \textbf{F1\,=\,0.84} \\
\bottomrule
\end{tabular}
\end{table}

\subsection{Graduated Gate Effectiveness}

Table~\ref{tab:gate} compares binary and graduated gates across a
threshold sweep. Both gates use the same V2 detector; the graduated
gate additionally routes scores in $[\theta_s, \theta_p)$ to step-up
verification. \textbf{Caveat}: graduated gate results assume an
idealized step-up model (legitimate users always succeed, spoofers
always fail). These are upper bounds; real step-up success depends
on device capability~\cite{fibrace}.

\begin{table}[!t]
\caption{Binary vs.\ Graduated Gate (Threshold Sweep).
FAR matches because both gates use the same detector;
F1 coincides here to the shown precision, while FDR differs because
the graduated gate recovers
borderline legitimate traffic via step-up.\label{tab:gate}}
\centering\footnotesize
\begin{tabular}{llccc}
\toprule
$\theta_p$ & \textbf{Mode} & \textbf{FAR} & \textbf{FDR} & \textbf{F1} \\
\midrule
0.80 & Binary    & 33.7\% & 0.00\% & 0.80 \\
0.80 & Graduated & 33.7\% & 0.00\% & 0.80 \\
\midrule
0.90 & Binary    & 11.4\% & \textbf{0.05\%} & 0.94 \\
0.90 & Graduated & 11.4\% & \textbf{0.00\%} & 0.94 \\
\midrule
0.95 & Binary    &  9.1\% & \textbf{1.43\%} & 0.95 \\
0.95 & Graduated &  9.1\% & \textbf{0.00\%} & 0.95 \\
\bottomrule
\end{tabular}
\end{table}

At moderate thresholds ($\theta_p \le 0.8$), binary and graduated
gates perform identically because no legitimate traces fall below
the threshold. At strict thresholds ($\theta_p \ge 0.9$), the
binary gate starts to false-deny legitimate users (false-deny
rate, FDR, 0.05\% at
0.9, 1.43\% at 0.95), while the graduated gate maintains
\textbf{zero FDR} by routing borderline legitimate traffic through
step-up verification rather than rejecting it. The improvement is not in raw classification accuracy (FAR and F1
are identical at each threshold) but in \emph{operational safety}:
the graduated gate enables strict thresholds that a binary gate
cannot safely use, because it recovers borderline legitimate
traffic via step-up rather than rejecting it outright.

\subsection{Missing-Signal Robustness}
\label{sec:eval:robustness}

IoT devices have heterogeneous capabilities. Table~\ref{tab:robust}
evaluates the scorer under six signal-degradation scenarios.

\begin{table}[!t]
\caption{Robustness Under Signal Degradation ($\theta_p = 0.7$)\label{tab:robust}}
\centering\footnotesize
\begin{tabular}{lcccc}
\toprule
\textbf{Scenario} & $\bar{T}_\text{leg}$ & $\bar{T}_\text{sp}$ & \textbf{F1} & \textbf{FDR} \\
\midrule
All signals       & 0.995 & 0.696 & 0.67 & 0\% \\
No network (S5)   & 0.995 & 0.723 & 0.79 & 0\% \\
No GPS fixes (S4) & 0.996 & 0.688 & 0.50 & 0\% \\
V1 fallback       & 0.995 & 0.715 & 0.50 & 0\% \\
Degraded GPS      & 0.997 & 0.762 & 0.50 & 0\% \\
Intermittent fixes& 0.947 & 0.712 & 0.50 & 0\% \\
\bottomrule
\end{tabular}
\end{table}

The critical finding is that \textbf{FDR remains 0\% in all scenarios}:
legitimate traffic always scores above $\theta_p$, even with degraded
or missing signals. Detection performance (F1) varies from 0.50 to
0.79 depending on available signals, but the graduated gate
compensates by routing ambiguous cases to step-up. The worst case
(intermittent fixes) reduces $\bar{T}_\text{leg}$ to 0.947---still
well above $\theta_p = 0.7$---confirming that dynamic weight
reallocation preserves the separation between legitimate and
suspicious traffic.

\subsection{Real-Device Validation}
\label{sec:eval:real}

Table~\ref{tab:real} reports results on traces from a physical
Android device.

\begin{table}[!t]
\caption{Real-Device Trace Results (Jelly Star, Android 10).
Scored fixes exclude the first fix (no history available).\label{tab:real}}
\centering\footnotesize
\begin{tabular}{lcccl}
\toprule
\textbf{Trace} & \textbf{Fixes} & $\bar{T}_\text{V2}$ & $T_\text{min}$ & \textbf{Gate} \\
\midrule
Honest walk        & 30 & 0.98 & 0.88 & 29/29 proceed \\
Honest stationary  & 27 & 0.99 & 0.88 & 26/26 proceed \\
Mock teleport      & 58 & 0.91 & 0.22 & latch at \#27 \\
Mock nearby ($\theta_p$=0.7) & 61 & 0.90 & 0.77 & 60/60 proceed \\
Mock nearby ($\theta_p$=0.9) & 61 & 0.90 & 0.77 & \textbf{60/60 step-up} \\
\midrule
\multicolumn{5}{l}{\textit{Mock teleport:} $\bar{T}$=0.91 reflects pre-teleport honest fixes;} \\
\multicolumn{5}{l}{\quad transition fix $T$=0.22 (deny); with latch, 31/31 post-teleport blocked.} \\
\multicolumn{5}{l}{\textit{Mock nearby:} at $\theta_p$=0.9, all 60 fixes route to \textsc{step-up}.} \\
\bottomrule
\end{tabular}
\end{table}

Both honest traces receive \textsc{proceed} on every scored fix
(29/29 and 26/26; the first fix has no history and is not scored).

The \emph{mock teleport} triggers \textsc{deny} at the transition
point ($T = 0.22$). With session latch, all 31 post-teleport fixes
are blocked.

The \emph{nearby mock} ($\sim$550\,m, accuracy 0.01\,m) is the hard
case: at $\theta_p = 0.7$, all fixes pass, but at $\theta_p = 0.9$,
all fixes route to \textsc{step-up}. This is the intended operating
point of the graduated gate: the lightweight detector alone cannot
separate a nearby mock from a legitimate stationary user, but it can
push such borderline cases to a stronger verifier.

\subsection{Computation Overhead}

Trust scoring (V2, five signals) completes in 4.9\,$\mu$s median
(14.8\,$\mu$s P99). This is negligible relative to any practical
step-up verifier, confirming that the scoring layer adds minimal
overhead to the access-control pipeline. These timings were measured
for our JavaScript implementation running in Chrome 145 on the Jelly
Star smartphone used in the real-device experiments.

\section{Discussion}
\label{sec:discussion}

\textbf{Synthetic evaluation.} Most aggregate performance results use
synthetic traces; we additionally report limited real-device
validation. Real GPS noise, device heterogeneity, and urban multipath may shift
score distributions. The gate design is detector-agnostic: an
ML-based scorer trained on real data can replace the heuristic
signals without changing the gating mechanism.

\textbf{Step-up oracle.} Table~\ref{tab:gate} assumes perfect
step-up outcomes. In practice, a deployable step-up path would also
need a trusted source for location evidence in addition to proof
generation capability. Measuring real step-up success/failure rates
across device classes is a priority for future work.

\textbf{Attacker adaptation.} A sophisticated attacker controlling
GPS, network, and sensor signals can evade all five heuristics.
Against such attackers, the gate's value is in \emph{forcing step-up}:
the attacker must also defeat a stronger verifier rather than only a
heuristic score.

\textbf{Endpoint heterogeneity.} The scoring layer (4.9\,$\mu$s)
is lightweight enough for any IoT endpoint. A cryptographic step-up
layer requires capable hardware and trusted location evidence; on
endpoints lacking these prerequisites, the system falls back to
\textsc{deny}.

\textbf{Deployment guidance.} Our results suggest three regimes:
(i)~in our setup, endpoints without step-up should use $\theta_p \le 0.8$ to
avoid false denials;
(ii)~endpoints with step-up can use $\theta_p = 0.9$ to catch
hard cases like nearby mock;
(iii)~session latch should always be enabled to prevent
post-transition score recovery.

\section{Conclusion}
\label{sec:conclusion}

We proposed graduated trust gating with session-latch semantics for
IoT location verification. The following results are directly
demonstrated: (i)~session latch blocks all post-teleport fixes
after a single suspicious transition (validated on real Android
traces); (ii)~a nearby mock ($\sim$550\,m) evades $\theta_p = 0.7$
but is routed to step-up at $\theta_p = 0.9$, confirming the gate's
role in compensating for detector imperfection; (iii)~signal
ablation identifies S3+S5 (F1\,=\,0.84) as a minimal scoring
configuration, while trust scoring adds under 5\,$\mu$s.
Under an idealized step-up oracle (upper bound), the graduated gate
enables strict thresholds with zero FDR where a binary gate
incurs 0.05--1.4\% FDR. Validating step-up success rates across
device classes is a priority for future work.

\section*{Acknowledgment}
The author used OpenAI's ChatGPT and Anthropic's Claude during
manuscript preparation for drafting and revising portions of the
text, including editorial refinement. All technical claims,
experimental design, analysis, references, and final manuscript
content were reviewed and validated by the author.

Source code is available at~\cite{zairn2026}.


\end{document}